# THE ROLE OF EXPANSION STRATEGIES AND OPERATIONAL ATTRIBUTES ON HOTEL PERFORMANCE: A COMPOSITIONAL APPROACH

Carles Mulet-Forteza, Berta Ferrer-Rosell,
Onofre Martorell Cunill and Salvador Linares-Mustarós

**Abstract**

**Purpose:** This study aims to explore the impact of expansion strategies and specific attributes of hotel establishments on the performance of international hotel chains, focusing on four key performance indicators: RevPAR, efficiency, occupancy, and asset turnover.

**Design/methodology/approach:** Data were collected from 255 hotels across various international hotel chains, providing a comprehensive assessment of how different expansion strategies and hotel attributes influence performance. The research employs compositional data analysis (CoDA) to address the methodological limitations of traditional financial ratios in statistical analysis.

**Findings:** The findings indicate that ownership-based expansion strategies result in higher operational performance, as measured by revenue per available room, but yield lower economic performance due to the high capital investment required. Non-ownership strategies, such as management contracts and franchising, show superior economic efficiency, offering more flexibility and reduced financial risk.

**Originality:** This study contributes to the hospitality management literature by applying CoDA, a novel methodological approach in this field, to examine the performance of different hotel expansion strategies with a sound and more appropriate method. The insights provided can guide hotel managers and investors in making informed decisions to optimize both operational and economic performance.

**Research limitations:** The study is limited to a dataset from 2019, and future research could extend the analysis to different time periods or consider the effects of external economic conditions, as well as other internal hotel/chain attributes.

**Practical implications:** The results provide actionable recommendations for hotel managers on selecting appropriate expansion strategies and optimizing hotel attributes to enhance performance.

**Social implications:** Understanding the performance impacts of different expansion strategies can contribute to sustainable business practices in the hospitality industry by promoting more efficient use of resources.

# 1. Introduction

Hotel chains can expand in various ways, with the most common being through growth strategies and diversification strategies (Martorell and Mulet, 2010). Diversification strategies involve developing new businesses, either related or unrelated to existing operations, such as investing in restaurant or laundry services. Growth strategies include franchise contracts, management contracts, ownership, leasing, or a combination of these (Martorell and Mulet, 2010). The relationship between these growth strategies and the performance they generate has received little attention from researchers in the hotel industry (Giannoukou, 2016).

Although growth is essential for any business, few studies examine the direct impact of growth strategies on hotel performance (Jang and Park, 2010; Giannoukou, 2016). This is largely due to the lack of publicly available data at the individual hotel level (Xiao *et al.*, 2012). This information gap has limited more detailed analysis. As most hotel managers consider the efficient selection of growth strategies crucial to improving hotel performance (Moon and Sharma, 2014), a disconnect between business practice and academic research becomes evident. Therefore, more research is needed to guide strategic decision-making for hotel managers (Giannoukou, 2016).

In general, the few studies conducted on this topic compare only a limited number of strategies, often without considering all alternatives or the broader chain perspective (Xiao *et al.*, 2012). In this context, this study seeks to provide empirical evidence on the following questions:

- Which growth strategies provide better operational and economic performance?
- What impact do related diversification activities, such as restaurant services and all-inclusive offerings, have on hotel performance?
- How do other hotel attributes, such as size, category, services offered, location, and seasonality, affect performance?

Regarding the methodological approach, given the large number of recent studies indicating the use of ratio-based variables in statistical analyses may produce inconsistent results (Creixans-Tenas *et al.*, 2019; Carreras-Simó *et al.*, 2021; Arimany-Serrat *et al.*, 2022; Coenders *et al.*, 2023a), we will use the compositional data analysis (CoDA) methodology (Aitchison, 1982). This methodology has been shown to be valid in methodological development studies for analyzing variables expressed as ratios (Pawlowsky-Glahn *et al.*, 2015; Linares-Mustarós *et al.*, 2018; 2022; Arimany-Serrat *et al.*, 2023; Molas-Colomer *et al.*, 2024), which are the types of data used in this study, as the performance measures (occupancy, revenue per available room (RevPAR), efficiency, and asset turnover) are expressed in ratios.

To address the questions raised, this study is structured as follows: The second section presents a literature review of studies that have analyzed the different growth strategies of hotel chains and their impact on performance. The following section defines the sample data, the statistical model variables, and the hypotheses to be tested. The statistical method used to answer the research questions is then described. Finally, the results obtained and the conclusions are presented.

## 2. Literature review

Regarding studies that have analyzed the relationship between growth strategies in the hotel industry and performance, the documents of Dev and Brown (1991), Brown and Dev (2000), Kim (2008), Hsu and Jang (2009), Moon and Sharma (2014), and Giannoukou (2016) stand out. Dev and Brown (1991) did not find significant differences in performance among the three growth strategies they analyzed: franchise, management, and ownership. In contrast, Brown and Dev (2000) found that hotel management contracts had a positive impact on performance. Furthermore, their study showed that hotels operated in the UK under management contracts were sold at a higher premium compared to those under lease contracts.

Kim (2008) examined the effects of management contracts on performance in the Korean hotel sector. The findings highlighted the positive effects of this type of contract on performance, particularly in terms of return on equity (ROE), liquidity, stability, and operating ratios. It was also shown that management contracts increased hotel competitiveness during tourism recessions. Overall, hotels with management contracts achieved higher performance compared to other models.

Similarly, Giannoukou (2016) analyzed a sample of luxury hotels in Greece, showing that the performance of hotels with management contracts was higher than that of franchised hotels, particularly in terms of ROS (retorn on sales), margin and turnover. These results also align with the research of Aissa and Goaied (2016), who stated that management contracts were associated with better performance. Finally, Hodari *et al.* (2017) showed that management contracts have a direct impact on the value of hotels, with long-term effects on assets and performance.

Hsu and Jang (2009) and Lin and Kim (2020) reached similar conclusions, although in these cases, a positive relationship between performance and franchise contracts was found. Mao and Mi's (2014) research revealed that hotels operated under franchise agreements achieved a higher RevPAR than those with management contracts. On the other hand, the study by Moon and Sharma (2014) evaluated franchise contracts from 21 publicly traded hotel chains in the United States and reached two main conclusions: first, they stated that chains with franchised hotels were more profitable than those without franchises; and second, they identified the optimal proportion of franchised and non-franchised hotels, determining that the optimal franchise proportion for maximizing return on assets was 76%. Finally, Lin and Kim's (2020) study highlighted that franchises offer greater flexibility and lower initial investment, while direct ownership provides greater control over service quality, which can improve operational performance.

Blal and Bianchi (2019) noted the lack of in-depth studies on the long-term effects of the non-equity growth model in the hotel industry, that is, those growth strategies not based on ownership, despite its growing popularity. In this context, Seo and Soh (2019) analyzed this model and discovered that investment sensitivities related to cash flows play a crucial role in determining ROE. The non-equity growth strategies (franchise, management, and lease) offer hotels greater financial flexibility by reducing costs associated with asset ownership and maximizing ROE, particularly in dynamic and changing environments.

Strouhal *et al.* (2018) emphasized how different methods for calculating earnings before interest and taxes significantly affect companies' financial performance. Märklin and Bianchi (2022) showed how the non-equity growth model influences the financial performance of hotel companies in complex environments. Similarly, Li and Singal (2019) and Seo *et al.* (2021) showed that the use of non-equity models allows for greater financial and operational flexibility, especially in rapidly changing markets.

In the hotel sector, few studies have examined the effect of diversification. Claver-Cortés *et al.* (2006) investigated diversification strategies in a sample of 80 Spanish hotels, but they did not reach conclusive results, similar to Bresciani *et al.* (2015). Chen and Chang (2012), in their analysis of Taiwanese hotels, found that those offering food and beverage services generated higher profits than those offering only accommodation, although with greater variability in performance. Mun *et al.* (2021) identified the food and beverage sector as a key component for differentiation and competitiveness in luxury hotels, particularly in Asia.

Similarly, Lee and Jang (2007) concluded that there is a positive relationship between related diversification and performance stability, but not necessarily with an increase in performance, findings also observed in López-Picos *et al.* (2017) for Spanish hotels. However, Mandić and Petrić (2021) pointed out that not all forms of diversification are equally effective; unrelated diversification can increase operational risks if synergies are not adequately managed.

On the other hand, many studies, lacking access to private data, have used public variables such as hotel size, category, location, or seasonality to determine hotel performance. For example, Claver-Cortés *et al.* (2007) analyzed the impact of hotel size and category on performance, concluding that the larger the hotel size and category, the greater its performance. Similarly, Bresciani *et al.* (2015) studied the relationship between hotel size and category and performance, finding a positive relationship between hotel category and performance, but no significant impact of size on performance.

## 3. Methodology

### 3.1. Study Data

The analysis of expansion strategies and hotel attributes in relation to performance was conducted using a sample of 255 hotels operating internationally and belonging to various hotel chains, whose identities will not be disclosed for confidentiality reasons. Unlike previous studies, which compare results at the corporate level, this study obtained data at the individual hotel level for the year 2019. The available information was provided directly by the hotel chains, allowing us to work with individualized and confidential data that are not available in any public database. The hotels operate in various international markets, with a majority presence in Europe and the Caribbean region. Furthermore, since these companies operate internationally, their results are of interest to many hotel group managers seeking strategies to improve performance.

The descriptive analysis shows that the performance of the hotels in the sample is heterogeneous, both operationally and financially. The average occupancy rate is 67.62%, although some hotels do not exceed 30%, while others operate at nearly 100% capacity. This variability is also reflected in other variables, with hotels showing high positive performance and others experiencing losses. It is also noteworthy that 7% of the hotels in the sample choose to close for nearly half of the year.

Additionally, RevPAR also varies significantly, ranging from €1.38 to €537.23, with an average of €107.78. As for economic performance, the average efficiency, defined as the ratio of revenue to expenses, remains around 200%, while the asset turnover ratio, defined as revenue to assets, stands at 63%, both figures being relatively high.

Regarding growth strategies, leasing is the most chosen option by hotel chains, accounting for 40% of all analyzed hotels. It is followed by management contracts, which account for 33%, and ownership, with 18%. Franchise contracts make up only 8% of the total. This shows that hotel chains prefer to expand through options that do not involve capital.

On the other hand, diversification among the hotels in the sample is also heterogeneous. On average, food and beverage (F&B) revenue represents 25% of total revenue, although there is significant variability. Some hotels are exclusively dedicated to accommodation, while in others, F&B accounts for up to 57% of revenue. It is also noteworthy that 18% of the hotels in the sample offer their services under an "all-inclusive" regime.

Regarding location, urban hotels account for slightly more than half (55%) of the hotels in the sample, compared to 45% for vacation hotels. This implies that a significant part of the supply is subject to a high level of seasonality. Most hotels are located in Europe (64%), while the Caribbean (16%) is the second most represented region in our sample.

In terms of size, when considering the number of employees, the average is 161, although in 41% of the hotels (those operated under management and franchise contracts), the staff is the responsibility of the owner rather than the operator. Some hotels have a substantial high number of employees, exceeding 1,500. This heterogeneity is also observed in the number of rooms, ranging from 19 to 1,176 rooms, with an average of 273 rooms per hotel. The heterogeneity in terms of seasonality and size is reflected in the wide range of sales volumes in our sample.

The quality of the hotels in the sample is high, with an average exceeding 4 stars. Table I summarizes the results of the descriptive analysis.

Table I. Descriptive analysis

| Variable | Mean | Std. Dev | Mín. | Max. |
|---|---|---|---|---|
| **PERFORMANCE** | | | | |
| Occupancy | 67.62 | 12.73 | 22.94 | 91.32 |
| RevPAR | 107.78 | 113.01 | 1.38 | 537.23 |
| Efficiency | 2.03 | 1.79 | 0.38 | 10.94 |
| Asset Turnover | 0.63 | 0.52 | 0.01 | 1.72 |
| **GROWTH STRATEGIES** | | | | |
| Leasing* | 0.40 | - | 0.00 | 1.00 |
| Management* | 0.33 | - | 0.00 | 1.00 |
| Franchise* | 0.08 | - | 0.00 | 1.00 |
| Ownership* | 0.18 | - | 0.00 | 1.00 |
| **DIVERSIFICATION** | | | | |
| % Food & Beverage | 0.25 | 0.13 | 0.00 | 0.57 |
| All-inclusive* | 0.18 | - | 0.00 | 1.00 |
| **LOCATION** | | | | |
| Urban* | 0.55 | - | 0.00 | 1.00 |
| Vacation* | 0.45 | - | 0.00 | 1.00 |
| Europe* | 0.64 | - | 0.00 | 1.00 |
| Caribbean* | 0.16 | - | 0.00 | 1.00 |
| Rest of the world* | 0.19 | - | 0.00 | 1.00 |
| **SIZE** | | | | |
| Employees | 161.01 | 198.06 | 2 | 1.541 |
| Rooms | 273.04 | 178.56 | 19 | 1.176.00 |
| Sales (thousands $) | 6.287 | 10.045 | 51 | 76.766 |
| **SEASONALITY** | | | | |
| Days open | 331.60 | 68.41 | 101 | 365 |
| **SERVICE QUALITY** | | | | |
| Stars | 4.20 | 0.61 | 2.00 | 5.00 |

*\* Dummy-coded qualitative variables. Mean values indicate proportions.*

## 3.2. Variables and Hypotheses

There is a general consensus that the measurement of performance should consider both operational and financial aspects (Bergin-Seers and Jago, 2007; Lado-Sestayo et al., 2017). In this study, we have considered four measures to evaluate the performance of each hotel at different levels. To formulate the dependent variables, we use the number of available rooms sold, the total number of hotel rooms, and total revenue.

Thus, to assess operational performance, we consider occupancy ($P_1$), defined as the proportion of rooms occupied to the total available rooms, and RevPAR ($P_2$), calculated by dividing total room revenue by the number of available rooms. We also use operating expenses and total assets to determine economic performance. In this case, we consider efficiency ($P_3$), measured as total revenue divided by operating expenses, and asset turnover ($P_4$), calculated as total revenue over total assets.

Regarding the effect that ownership may have on performance, we consider that, at an operational level, owned hotels could achieve better results. This is because hotel chains tend to select preferred locations for the hotels where they commit their resources, and ownership provides greater control over operations, resulting in better service quality (Contractor and Kundu, 1998a. Owned assets could have differentiating characteristics that provide the company with a competitive advantage over other options.

According to the resource theory and the theory of transaction cost, as pointed out by Kruesi et al. (2018), ownership allows for better preservation of intangibles and specific assets, which can constitute a competitive advantage compared to alternatives that do not involve ownership. Since it is logical to opt for control when assets offer a competitive advantage, we believe that hotel chains purchase assets when they detect specific and differentiating characteristics. These hotels may also better align with management goals in terms of design, efficiency, and location. Anderson and Gatignon (1986) highlight that ownership is one of the most effective options for maintaining product and service quality, avoiding issues related to information loss and lack of franchisee involvement. Therefore, we consider that owned hotels may achieve better operational results in terms of RevPAR and occupancy.

At the economic level, owned hotels may present lower performance ratios compared to other strategies that do not involve capital, as real estate investment increases the total value of assets. An exception could be leasing, although this strategy has the disadvantage of requiring the payment of a fixed cost, usually corresponding to high rents (Kruesi et al., 2018). This fact had significant consequences during the financial crisis at the end of the first decade of the 21st century, when hotels continued paying high rents even though they were generating little to no revenue (Kruesi et al., 2018). Based on this, we hypothesize that economic performance may be lower for hotels operated through an ownership strategy.

In summary, the first two hypotheses are as follows:

H1. Owned hotels present higher operational performance than those operated through other expansion strategies.

H2. Owned hotels present lower economic performance than those operated through other expansion strategies.

The relationship between diversification and performance has been studied in various sectors without reaching definitive conclusions. Theoretically, sharing resources among different business units could increase performance and reduce risk, although additional costs generated at very high levels of diversification could counteract the synergies produced (Hitt *et al.*, 1997; Nickel and Rodríguez, 2002). In the hotel sector, research on this topic is limited, and the results are mixed. Some studies (Lee and Jang, 2007; Chen and Chang, 2012) support a positive effect on performance, while others suggest an improvement in the stability of profits but no significant increase in performance (Lee and Jang, 2007). Based on theoretical and empirical foundations, we propose the following two hypotheses regarding related diversification:

H3. Related diversification has a positive effect on hotel performance (the percentage of revenue from food and beverage is used as an indicator of related diversification). We considered that a hotel offers restaurant services if more than 20% of the hotel's revenue comes from food and beverage services.

H4. Hotels offering an all-inclusive service present better performance.

The hotel size can influence its performance. Several authors (Claver-Cortés *et al.*, 2007; Bresciani *et al.*, 2015) have established a positive relationship between size and performance, based on the existence of economies of scale. However, other authors suggest that this relationship may not be linear, as the positive effect of economies of scale could be offset by higher exit barriers due to a larger hotel size (Lado-Sestayo *et al.*, 2017; Bresciani *et al.*, 2015). Based on this, we aim to test the following hypothesis:

H5. The relationship between size and performance is positive (the natural logarithm of the number of employees at the hotel is used as an indicator of size).

The number of stars a hotel has is an indicator of quality and services offered to guests. Several previous studies have found a positive relationship between the number of stars and performance, suggesting that the higher service quality associated with hotels with more stars leads to greater performance. Studies such as those by Brown and Dev (1999), Bresciani *et al.* (2015), Pine and Phillips (2005), and Claver-Cortés *et al.* (2007) support this positive relationship between a hotel's category and its performance. Therefore, the hypothesis to be tested is as follows:

H6. The number of stars a hotel has has a positive effect on its performance.

Seasonality is a significant factor in the hotel sector and has been examined by various authors in the academic literature (Spencer and Holecek, 2007; Cardona, 2014). It has been observed that all tourist destinations experience some degree of seasonality, more pronounced in coastal (Karyopouli and Koutra, 2012; Koutra and Karyopouli, 2013) than in urban destinations (Martín *et al.*, 2014). Additionally, the location of a hotel influences its performance (Sainaghi, 2011), with highly seasonal regions are more at risk due to the fluctuation in tourist income. As indicators of seasonality, the number of days the hotel is open and whether it is located in an urban or vacation area are used. In this context, the following two hypotheses are proposed:

H7. Hotels located in urban areas have higher performance.

H8. Seasonality (measured by the number of days hotels are open) negatively affects performance. A hotel is considered seasonal if it has been open for less than 9 months of the year.

Table II shows the independent variables, hypotheses, and expected results regarding performance, as previously discussed.

Table II. Independent variables, hypotheses, and expected sign.

| Variable | Definition | Hypothesis | Operational performance | Economic performance |
|---|---|---|---|---|
| Leasing | Dummy that takes value 1 when the hotel is in Leasing | H1, H2 | - | + |
| Management | Dummy that takes value 1 when the hotel is in a management contract | H1, H2 | - | + |
| Franchise | Dummy that takes value 1 when the hotel is in franchise | H1, H2 | - | + |
| Restaurant | Food & beverage revenues / Total revenue | H3 | + | + |
| All-inclusive | Dummy that takes value 1 when the hotel is all-inclusive | H4 | + | + |
| Size | Logarithm of number of employees | H5 | + | + |
| Stars | Number of stars | H6 | + | + |
| Location [Urban] | Dummy that takes value 1 when the hotel is urban | H7 | + | + |
| Seasonal [days opent] | Number of days open | H8 | + | + |

### 3.3. Method

The statistical analysis of the data was conducted using the CoDA methodology. Created by Aitchison (1982), CoDA has a solid mathematical modeling foundation (Pawlowsky-Glahn *et al.*, 2015), which has established it as one of the most reliable methodologies for handling variables expressed as ratios (Coenders and Ferrer-Rosell, 2020; Coenders *et al.*, 2023b).

Although CoDA offers various options for addressing problems where ratios are an essential part of the study, such as additive log-ratios (alr), center log-ratios (clr), or isometric log-ratios (ilr), this study opted for the use of pairwise log-ratios (Coenders and Arimany-Serrat, 2023; Hron *et al.*, 2021; Creixans-Tenas *et al.*, 2019; Greenacre, 2019). Since these ratios are similar to applying a logarithm to classic ratios, where there is only one component in the numerator and another in the denominator, their use facilitates the acceptance of the CoDA method as it is perceived as a simple logarithmic scale change.

In addition, to ensuring the validity of the results, using CoDA with pairwise log-ratios offers several advantages. First, these transformations create variables within the range $(-\infty, +\infty)$, consistent with what is expected from variables that follow a normal distribution (Linares-Mustarós *et al.*, 2018). It is important to recall that the assumption of normality of the variables is necessary in linear regression studies, as is the case in this study.

Secondly, it is important to note that the results of linear regression models using ratios as variables are often inconsistent if the numerator and denominator of these ratios are permuted. This methodological issue is resolved through the use of log-ratios (Coenders *et al.*, 2023a; Linares-Mustarós *et al.*, 2022; Molas-Colomer *et al.*, 2024). Additionally, CoDA transformations reduce the non-linearity of the variables in linear regression models (Carreras-Simó and Coenders, 2021), thus improving the linearity assumption of the variables, which is essential in any linear regression study. It is worth noting that traditional accounting ratios, such as efficiency and asset turnover, do not meet this assumption (Carreras-Simó and Coenders, 2021).

Thirdly, thanks to the construction of a connected acyclic graph based on the variables used to create the pairwise log-ratios, it is easier in CoDA to identify whether there are redundant variables in the model. This avoids including variables that can be derived from a linear relationship with other variables in the model (Creixans-Tenas *et al.*, 2019; Greenacre, 2019).

Once the methodology has been justified, the regression model used is presented, along with the independent and dependent variables and the graph that provides the CoDA ratios. A multivariate linear regression model is applied to analyze the data in this study. The performance ratios acting as dependent variables in the model are, for operational performance, the logarithmic transformation of occupancy ($P_1$) and RevPAR ($P_2$). For economic performance, the dependent variables are the logarithmic transformations of efficiency ($P_3$) and asset turnover ($P_4$).

The pairwise log-ratios are used as independent variables in the regression model and are defined as follows:

ln($P_1$) = ln (occupied rooms/available rooms)
ln($P_2$) = ln (revenue/occupied rooms)
ln($P_3$) = ln (revenue/expenses)
ln($P_4$) = ln (revenue/assets)

and correspond to the connected acyclic graph in Figure 1.

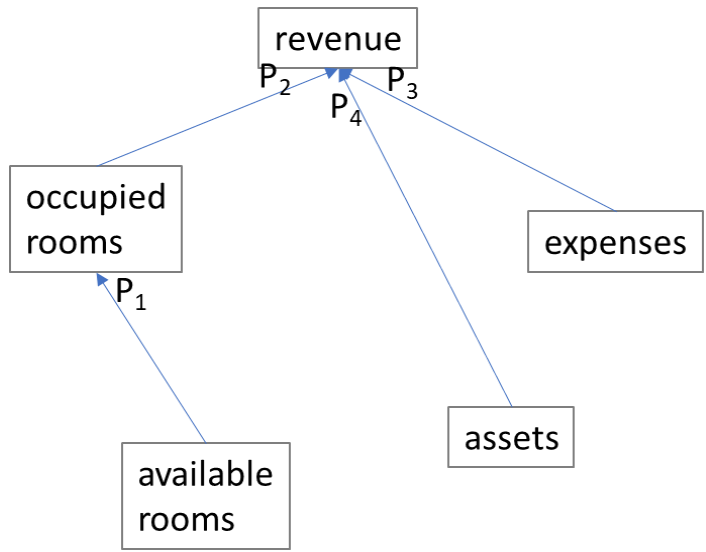

Figure 1. Connected Acyclic Graph.

Following Greenacre (2019), the arrows in the graph indicate which variables are in the numerator (arrowhead) and which are in the denominator (arrow tail) of the CoDA ratios. The inability to reach one element of the graph from another through two different paths ensures the linear independence of the dependent variables, thus avoiding redundancy in the dependent variables.

On the other hand, the independent variables considered in this study are as follows: lease, management, franchise, restaurant, all-inclusive, size (natural logarithm of the number of employees), stars, urban location, seasonality, Caribbean, and rest of the world (RM).

Based on all the above, the proposed multivariate regression model is as follows:

$\ln(P_i) = \alpha + \beta_1$ Leasing $+ \beta_2$ Managment $+ \beta_3$ Franchise $+ \beta_4$ Restaurant $+$
$+ \beta_5$ All-incluisve $+ \beta_6$ Size $+ \beta_7$ Stars $\beta_8$ Urban $+ \beta_9$ Seasonal
$+ \beta_{10}$ Caribbean $+ \beta_{11}$ Rest of the world $+ \varepsilon_i$

Where $\ln(P_i)$ is the natural logarithm of the performance ratio used (Occupancy, RevPAR, Efficiency, and Asset Turnover).

## 4. Results

Table III shows the results obtained from the multivariate regression model.

Table III. Results of the multivariate linear regression model

| Variable | Occupancy Ln($P_1$) | RevPAR Ln($P_2$) | Efficiency Ln($P_3$) | Asset Turnover Ln($P_4$) |
|---|---|---|---|---|
| Leasing | 0.017 | -0.141 * | -0.218 ** | -0.119 |
| Management | -0.031 | -3.157 *** | 0.807 *** | -2.578 *** |
| Franchise | -0.304 *** | -3.033 *** | 1.353 *** | -2.731 *** |
| Restaurant | -0.095 *** | -0.199 *** | -0.213 *** | -0.260 ** |
| All-inclusive | 0.069 | -0.187 * | -0.171 | -0.327 * |
| Size | 0.059 *** | 0.182 *** | 0.158 *** | -0.246 *** |
| Stars | -0.018 | 0.448 *** | 0.136 ** | 0.470 *** |
| Urban | -0.113 ** | -0.062 | -0.150 * | -0.328 ** |
| Seasonal | -0.081 * | -0.379 *** | -0.142 | -0.670 *** |
| Caribbean | -0.315 *** | -0.265 * | -0.319 ** | -0.259 |
| Rest of the world | -0.120 ** | -0.245 ** | -0.260 ** | -0.110 |
| $R^2$ | 0.382 | 0.947 | 0.675 | 0.865 |

*Notes: * significant at 5%; ** significant at 1%; *** significant at 0.1%.*

As can be observed, growth strategies based on ownership contracts achieve higher operational performance, measured by RevPAR, compared to non-ownership strategies (a negative effect is detected for all three non-equity strategies). In the case of occupancy, this significant difference is only observed in comparison with franchising. Therefore, the results support Hypothesis 1, which posits that owned hotels have better operational performance due to greater control over operations, higher service quality (Contractor and Kundu, 1998a, and their unique characteristics (location, building type, design, service, facilities, etc.), which provide a competitive advantage over other alternatives.

Our study supports the theoretical framework of resource theory and theory of transaction costs, which argue that ownership better preserves intangibles and specific assets, providing a competitive advantage compared to non-ownership alternatives (Kruesi *et al.*, 2018). Additionally, our results align with the theory that hotel chains acquire assets when they identify specific and differentiating characteristics.

On the other hand, franchise and management contracts have a positive effect on the economic performance of hotels, considering only efficiency (P3), partially confirming Hypothesis 2, which assumed that the high investment would limit the economic performance of hotels managed under ownership contracts. However, regarding asset turnover, the opposite of what was expected is observed, as hotels managed through ownership contracts generate sufficient revenue to offset the higher investment made.

In the case of hotels with lease contracts, efficciency is negatively affected, consistent with Kruesi *et al.* (2018). This result is primarily due to the high fixed cost of rent, which must be paid by the managers of leased hotels, ultimately generating a negative effect on margin.

Regarding diversification strategies, the results show that related diversification does not always lead to improved hotel performance. Specifically, we find that having a restaurant, while relevant for a hotel, significantly harms all performance measures, as it does not compensate for the costs associated with this infrastructure.

Therefore, our results contradict Hypothesis 3, which justified diversification based on synergies and the sharing of resources and knowledge between business units (Hitt *et al.*, 1997; Nickel and Rodríguez, 2002). However, these results partially support the conclusions of Chen and Chang (2012), who pointed out that diversified companies incur higher levels of risk. Additionally, these results are consistent with those for the 'all-inclusive' variable, as this service worsens both RevPAR and asset turnover, also contradicting Hypothesis 4.

Regarding hotel size, this variable is significant in all the equations considered and is relevant in terms of both operational and economic performance. In this sense, our study supports Hypothesis 5, as hotel size provides significant economies of scale in terms of better occupancy, RevPAR and efficiency, a result consistent with previous studies such as those by Pine and Phillips (2005) and Claver-Cortés *et al.* (2007).

As for the number of stars, we observe that this variable is significant for three of the four performance measures, indicating that higher quality leads to better results in terms of RevPAR, efficiency, and asset turnover. Therefore, our results support Hypothesis 6, which posits a positive relationship between hotel quality and performance, as also demonstrated in previous studies such as Brown and Dev (1999), Pine and Phillips (2005), and Claver-Cortés *et al.* (2007).

It is also interesting to note that urban hotels show a reduction in occupancy, efficiency, and asset turnover levels. Our results contradict Hypothesis 7, which expected a positive relationship with all performance indicators, in line with what has been proposed in some previous studies (Karyopouli and Koutra, 2012; Koutra and Karyopouli, 2013; Martín *et al.*, 2014). Therefore, our study questions the association between lower seasonality and urban hotels, as indicated in earlier studies.

We consider that our study results may be influenced by the fact that many of the hotels in our database operate in warm-weather destinations all year round, such as the Caribbean and the Canary Islands, which have even lower seasonality than urban hotels located in inland cities. This could be explained by the fact that urban hotels, especially those focused on business tourism, tend to have lower occupancy during weekends and the month of August. We also observe that hotel seasonality is negatively related to RevPAR, occupancy, and asset turnover, thus confirming the last hypothesis of our study.

Furthermore, it is demonstrated that hotels located in the Caribbean or other regions outside of Europe show lower levels of RevPAR, occupancy, and efficiency compared to hotels located in Europe. This could be due to the fact that Europe is still perceived as a prominent destination, where services such as infrastructure, health, safety, and economic stability can be better monetized compared to other regions.

## 5. Conclusions

This study has analyzed the impact of expansion strategies and individual hotel attributes on performance in international hotel chains, providing insights into how these factors influence hotel performance. The results are valuable for both investors and hotel operators. They enable investors to assess which expansion strategies and hotel characteristics lead to greater operational and economic performance, facilitating informed investment decisions. For managers, the findings highlight the importance of selecting appropriate growth and diversification strategies and optimizing hotel attributes, such as size, location, category, and services offered, to maximize operational and economic efficiency.

The results obtained have significant implications from both a theoretical and practical perspective. Firstly, regarding the practical implications for the international hotel industry, it is confirmed that ownership strategies tend to offer higher operational performance, especially in terms of RevPAR. This suggests that hotel chains seeking to maximize revenue should consider maintaining ownership of their establishments, particularly in strategic locations or those with unique characteristics that provide a competitive advantage (Contractor and Kundu, 1998b). However, this strategy also entails lower economic performance, primarily due to the high initial investment costs associated with it (Li and Singal, 2019), implying the need to optimize operating expenses and adjust cost structures to counterbalance these investments.

In this context, the growth model based on non-equity strategies has proven to be a more flexible and less risky option in dynamic environments (Blal and Bianchi, 2019; Märklin and Bianchi, 2022). From a management perspective, the results indicate that hotel chains must carefully analyze the different available growth strategies (franchise, management, lease, ownership) and how they affect performance in both the short and long term.

Additionally, diversification into services such as food and beverage or all-inclusive packages, while it may contribute to differentiation, must be adequately scaled to avoid becoming a cost burden that does not generate sufficient returns (Mun *et al.*, 2021; Mandić and Petrić, 2021). This implies that managers must be very mindful of operational efficiency and adjust the service offering to real demand, while keeping customer satisfaction as a key determinant of performance (Nazari *et al.*, 2020).

Secondly, regarding theoretical implications, this study provides new evidence for the resource theory and theory of transaction costs, confirming that owned assets can offer competitive advantages in terms of operational performance due to their location and unique characteristics (Kruesi *et al.*, 2018). It also reinforces the idea that related diversification strategies are not always beneficial, especially when the associated costs outweigh the expected synergies (Hitt *et al.*, 1997).

Furthermore, the assumption that urban hotels always have better performance has been questioned, demonstrating that seasonality and specific location can negatively influence hotel performance (Karyopouli and Koutra, 2012; Koutra and Karyopouli, 2013).

This study presents some limitations that should be considered when interpreting the results. Firstly, data availability and accessibility represent a significant restriction. The sample includes 255 hotels, but the data collected is specific to the year 2019 and comes from a single restricted-access database, which may limit the generalizability of the results. However, the use of 2019 data also presents important advantages, as it avoids the distortion caused by the impact of the COVID-19 pandemic. Thus, the results are valid in the current context, where the influence of the pandemic is minimal and tourism indicators have recovered or even surpassed 2019 levels. In this way, the results reflect a more representative situation of typical tourism dynamics.

Secondly, regarding the methodology, the use of CoDA represents a significant advance, but its application is completely new in the field of hotel management, and relatively new in hospitality management (Saus-Sala *et al.*, 2024) This may complicate comparability with previous studies using traditional methods, even though pairwise log-ratios to measure performance are straightforward to intèrpret.

Finally, another limitation is geographical diversity and seasonality. Although the hotels in the sample operate in different regions, the analysis does not explore how regional differences, such as tourism demand or local policies, may affect performance. Additionally, seasonality is considered in a simplified manner, which may not fully capture the complexity of this phenomenon in different markets.

Regarding hotel attributes, they have been treated in a simplified manner, such as measuring diversification only by the percentage of revenue from food and beverage without considering other sources of diversification. This may limit a complete understanding of the impact of different aspects of diversification on performance. The study focuses on four main growth strategies (franchise, management, lease, and ownership), leaving out other hybrid forms that may exist. This may restrict the scope of the conclusions, especially in such a dynamic industry as the hotel industry, where strategies can be more complex and varied.

Lastly, the performance measures used in the study, based on RevPAR, occupancy, efficiency, and asset turnover, do not include other factors that could also significantly influence long-term performance, such as customer satisfaction, loyalty, or environmental impact. This implies that the analysis focuses exclusively on economic and operational aspects, without considering other dimensions that could provide a more holistic view of hotel performance.

Overall, this study provides a solid foundation for decision-making in hotel management, offering guidance for optimizing growth strategies and configuring hotel attributes with the goal of maximizing both operational and economic performance in a highly competitive environment, using a more suitable methodological tool.

**References**

Aissa, S.B. and Goaied, M. (2016), "Determinants of Tunisian hotel profitability: The role of managerial efficiency", *Tourism Management*, Vol. 52, pp. 478-487. https://doi.org/10.1016/j.tourman.2015.07.015

Aitchison, J. (1982), "The statistical analysis of compositional data (with discussion)", *Journal of the Royal Statistical Society Series B (Statistical Methodology)*, Vol. 44 No. 2, pp. https://doi.org/10.1111/j.2517-6161.1982.tb01195.x

Anderson, E. and Gatignon, H. (1986), "Modes of foreign entry: a transaction cost analysis and propositions", *Journal of International Business Studies*, Vol. 17 No. 3, pp. 1-26. https://doi.org/10.1057/palgrave.jibs.8490432

Arimany-Serrat, N., Farreras-Noguer, M.À. and Coenders, G. (2022), "New developments in financial statement analysis. Liquidity in the winery sector", *Accounting*, Vol. 8, pp. 355-366. https://doi.org/10.5267/j.ac.2021.10.002

Arimany-Serrat, N., Farreras-Noguer, M.À. and Coenders, G. (2023), "Financial resilience of Spanish wineries during the COVID-19 lockdown", *International Journal of Wine Business Research*, Vol. 35 No. 2, pp. 346-364. https://doi.org/10.1108/IJWBR-03-2022-0012

Bergin-Seers, S. and Jago, L. (2007), "Performance measurement in small motels in Australia", *Tourism and Hospitality Research*, Vol. 37 No. 2, pp. 144-155.

Blal, I. and Bianchi, G. (2019), "The asset-light model: A blind spot in hospitality research", *International Journal of Hospitality Management*, Vol. 76, pp. 39-42. https://doi.org/10.1016/j.ijhm.2018.02.021

Bresciani, S., Thrassou, A. and Vrontis, D. (2015), "Determinants of performance in the hotel industry - an empirical analysis of Italy", *Global Business and Economics Review*, Vol. 17 No. 1, pp. 19-34. https://doi.org/10.1504/GBER.2015.066531

Brown, J.R. and Dev, C.S. (1999), "Looking beyond RevPAR", *Cornell Hotel and Restaurant Administration Quarterly*, Vol. 40 No. 2, pp. 23-33.

Brown, J. and Dev, C. (2000), "Improving productivity in a service business: Evidence from the hotel industry", *Journal of Service Research*, Vol. 2 No. 4, pp. 339-354.

Cardona, J.R. (2014), "La estacionalidad turística e sus potenciales impactos", *Rosa dos Ventos*, Vol. 6 No. 3, pp. 446-468.

Carreras-Simó, M. and Coenders, G. (2021), "The relationship between asset and capital structure: compositional approach with panel vector autoregressive models", *Quantitative Finance and Economics*, Vol. 5, pp. 571-590. doi: 10.3934/QFE.2021025

Chen, C. and Chang, K. (2012), "Diversification strategy and financial performance in the Taiwanese hotel industry", *International Journal of Hospitality Management*, Vol. 31, pp. 1030-1032. https://doi.org/10.1016/j.ijhm.2011.10.003

Claver-Cortés, E.C., Guerrero, R.A. and Ramón, D.Q. (2006), "Las ventajas de la diversificación estratégica para las empresas turísticas españolas", *Cuadernos de Turismo*, Vol. 17, pp. 51-74.

Claver-Cortés, E., Pereira, J. and Molina, J. (2007), "Impacto del tamaño, el tipo de gestión y la categoría sobre el desempeño de los hoteles españoles", *Cuadernos de Turismo*, Vol. 19, pp. 27-45.

Coenders, G. and Arimany-Serrat, N. (2023), "Accounting statement analysis at industry level. A gentle introduction to the compositional approach", *arXiv*. Available at: https://arxiv.org/abs/2305.16842.

Coenders, G., Sgorla, A.F., Arimany-Serrat, N., Linares-Mustarós, S. and Farreras-Nogueras, M.A. (2023a), "Nous mètodes estadístics composicionals per a l'anàlisi de ràtios comptables", *Revista de Comptabilitat i Direcció*, Vol. 35, pp. 133-146.

Coenders, G., Egozcue, J.J., Fačevicová, K., Navarro-López, C., Palarea-Albaladejo, J., Pawlowsky-Glahn, V. and Tolosana-Delgado, R. (2023b), "40 years after Aitchison's article "The statistical analysis of compositional data". Where we are and where we are heading", *SORT. Statistics and Operations Research Transactions*, Vol. 47 No. 2, pp. 207–228. DOI: 10.57645/20.8080.02.6

Coenders, G. and Ferrer-Rosell, B. (2020), "Compositional data analysis in tourism. Review and future directions", *Tourism Analysis*, Vol. 25 No. 1, pp. 153-168. https://doi.org/10.3727/108354220X15758301241594

Contractor, F.J. and Kundu, S.K. (1998a), "Modal choice in a world of alliances: analyzing organizational forms in the international hotel sector", *Journal of International Business Studies*, Vol. 29, pp. 325-358.

Contractor, F.J. and Kundu, S.K. (1998b), "Franchising versus company run operations: modal choice in the global hotel sector", *Journal of International Marketing*, Vol. 6 No. 2, pp. 28-53.

Creixans-Tenas, J., Coenders, G. and Arimany-Serrat, N. (2019), "Corporate social responsibility and financial profile of Spanish private hospitals", *Heliyon*, Vol. 5, Article e02623. DOI: 10.1016/j.heliyon.2019.e02623

Dev, C. and Brown, J. (1991), "Franchising and other operating arrangements in the lodging industry: A strategic comparison", *Hospitality Research Journal*, Vol. 14 No. 3, pp. 23-41. https://doi.org/10.1177/109634809101400

Giannoukou, I. (2016), "Strategic development and operation of deluxe hotels operating in Greece: management contracts vs franchising", *Journal of Organisational Studies and Innovation*, Vol. 3 No. 4, pp. 31-50.

Greenacre, M. (2019), "Variable selection in compositional data analysis using pairwise logratios", *Mathematical Geosciences*, Vol. 51 No. 5, pp. 649-682. https://doi.org/10.1007/s11004-018-9754-x

Hitt, M.A., Robert, H. and Hicheon, K. (1997), "International diversification: Effects on innovation and firm performance in product-diversified firms", *Academy of Management Journal*, Vol. 40 No. 4, pp. 767-798. https://doi.org/10.5465/256948

Hodari, D., Balla, P.J. and Aroul, R.R. (2017), "The matter of encumbrance: How management structure affects hotel value", *Cornell Hospitality Quarterly*, Vol. 58 No. 3, pp. 293-311. https://doi.org/10.1177/19389655166861

Hron, K., Coenders, G., Filzmoser, P., Palarea-Albaladejo, J., Faměra, M. and Matys-Grygar, T. (2021), "Analysing pairwise logratios revisited", *Mathematical Geosciences*, Vol. 53 No. 7, pp. 1643-1666. https://doi.org/10.1007/s11004-021-09938-w

Hsu, L. and Jang, S. (2009), "Effects of restaurant franchising: does an optimal franchise proportion exist?", *International Journal of Hospitality Management*, Vol. 28, pp. 204-211. https://doi.org/10.1016/j.ijhm.2008.07.002

Jang, S. and Park, K. (2010), "Hospitality finance research during recent two decades: subject, methodology, and citations", *International Journal of Contemporary Hospitality Management*, Vol. 23 No. 4, pp. 479-497. https://doi.org/10.1108/09596111111129995

Karyopouli, S. and Koutra, C. (2012), "Cyprus as a winter destination: An exploratory study", *Tourism Analysis*, Vol. 17 No. 4, pp. 495-508. https://doi.org/10.3727/108354212X13473157390803

Kim, S. (2008), "Hotel management contract: impact on performance in the Korean hotel sector", *The Service Industries Journal*, Vol. 28 No. 5, pp. 710-718. https://doi.org/10.1080/02642060801988332

Koutra, C. and Karyopouli, S. (2013), "Cyprus' image-a sun and sea destination- as a detrimental factor to seasonal fluctuations. Exploration into motivational factors for holidaying in Cyprus", *Journal of Travel and Tourism Marketing*, Vol. 30 No. 7, pp. 700-714. https://doi.org/10.1080/10548408.2013.827548

Kruesi, M.A., Hemmington, N.R. and Kim, P.B. (2018), "What matters for hotel executives? An examination of major theories in non-equity entry mode research", *International Journal of Hospitality Management*, Vol. 70, pp. 25-36. https://doi.org/10.1016/j.ijhm.2017.11.005

Lado-Sestayo, R., Vivel-Bua, M. and Otero-González, L. (2017), "Determinants of TRevPAR: hotel, management and tourist destination", *International Journal of Contemporary Hospitality Management*, Vol. 29 No. 12, pp. 3138-3156. https://doi.org/10.1108/IJCHM-03-2016-0151

Lee, M.J. and Jang, S.S. (2007), "Market diversification and financial performance and stability: A study of hotel companies", *International Journal of Hospitality Management*, Vol. 2, pp. 362-375. https://doi.org/10.1016/j.ijhm.2006.02.002

Li, Y. and Singal, M. (2019), "Capital structure in the hospitality industry: The role of the asset-light and fee-oriented strategy", *Tourism Management*, Vol. 70, pp. 124-133. https://doi.org/10.1016/j.tourman.2018.08.004

Lin, S.C. and Kim, Y.R. (2020), "Diversification strategies and failure rates in the Texas lodging industry: Franchised versus company-operated hotels", *International Journal of Hospitality Management*, Vol. 88, Article 102525. https://doi.org/10.1016/j.ijhm.2020.102525

Linares-Mustarós, S., Coenders, G. and Vives-Maestres, M. (2018), "Financial performance and distress profiles. From classification according to financial ratios to compositional classification", *Advances in Accounting*, Vol. 40, pp. 1-10. https://doi.org/10.1016/j.adiac.2017.10.003

Linares-Mustarós, S., Farreras-Noguer, M.À., Arimany-Serrat, N. and Coenders, G. (2022), "New financial ratios based on the compositional data methodology", *Axioms*, Vol. 11, Article 694. https://doi.org/10.3390/axioms11120694

López-Picos, Y., Otero-González, L. and Lado-Sestayo, R. (2017), "Efectos de la diversificación en el binomio de rentabilidad-riesgo. Un análisis del sector hotelero", *Revista de Investigaciones Turísticas*, Vol. 16, pp. 3-22.

Mandić, A. and Petrić, L. (2021), "The impacts of location and attributes of protected natural areas on hotel prices, implications for sustainable tourism development", *Environment, Development and Sustainability*, Vol. 23 No. 1, pp. 833-863. https://doi.org/10.1007/s10668-020-00611-6

Mao, Z. and Mi, Q. (2014), "An investigation of the relationship between selected hotel characteristics and performance in the extended stay hotel segment", *Journal of Tourism and Hospitality Management*, Vol. 2 No. 5, pp. 212-222. DOI: 10.17265/2328-2169/2014.05.003

Märklin, P. and Bianchi, G. (2022), "A differentiated approach to the asset-light model in the hotel industry", *Cornell Hospitality Quarterly*, Vol. 63 No. 3, pp. 313-319. https://doi.org/10.1177/19389655211006076

Martín, J.M., Jiménez, J. and Molina, V. (2014), "Impacts of seasonality on environmental sustainability in the tourism sector based on destination type: an application to Spain's Andalusia region", *Tourism Economics*, Vol. 20 No. 1, pp. 123-142. https://doi.org/10.5367/te.2013.0256

Martorell, O. and Mulet, C. (2010), "The franchise contract in hotels chains: a study of hotel chain growth and market concentrations", *Tourism Economics*, Vol. 16 No. 3, pp. 493-515. https://doi.org/10.5367/000000010792278446

Molas-Colomer, X., Linares-Mustarós, S., Farreras-Noguer, M.À. and Ferrer-Comalat, J.C. (2024), "A new methodological proposal for classifying firms according to the similarity of their financial structures based on combining compositional data with fuzzy clustering", *Journal of Multiple-Valued Logic and Soft Computing*, Vol. 43 Nos 1-2, pp. 73-100.

Moon, J. and Sharma, A. (2014), "Franchising effects on the lodging industry: optimal franchising proportion in terms of profitability and intangible value", *Tourism Economics*, Vol. 20 No. 4, pp. 695-725. https://doi.org/10.5367/te.2013.0336

Mun, S.G., Woo, L. and Seo, K. (2021), "Importance of F&B operation in luxury hotels: the case of Asia versus the US", *International Journal of Contemporary Hospitality Management*, Vol. 33 No. 1, pp. 125-144. https://doi.org/10.1108/IJCHM-06-2020-0546

Nazari, N., Abd Rahman, A. and Ab Aziz, Y.B. (2020), "The effect of customer satisfaction on the performance of the small and medium-sized hotels", *Tourism and Hospitality Management*, Vol. 26 No. 1, pp. 69-96.

Nickel, M.N. and Rodríguez, M.C. (2002), "A review of research on the negative accounting relationship between risk and return: Bowman's paradox", *Omega*, Vol. 1, pp. 1-18. https://doi.org/10.1016/S0305-0483(01)00055-X

Okumus, F. (2002), "Can hospitality researchers contribute to the strategic management literature?", *International Journal of Hospitality Management*, Vol. 21 No. 2, pp. 105-110. https://doi.org/10.1016/S0278-4319(01)00033-0

Pawlowsky-Glahn, V., Egozcue, J.J. and Tolosana-Delgado, R. (2015), *Modeling and analysis of compositional data*, Wiley, Chichester.

Pine, R. and Phillips, P.A. (2005), "Performance comparisons of hotels in China", *International Journal of Hospitality Management*, Vol. 24 No. 1, pp. 7-73. https://doi.org/10.1016/j.ijhm.2004.04.004

Sainaghi, R. (2011), "RevPAR determinants of individual hotels: evidences from Milan", *International Journal of Contemporary Hospitality Management*, Vol. 23 No. 3, pp. 297-311. https://doi.org/10.1108/09596111111122497

Saus-Sala, E., Farreras-Noguer, M.À., Arimany-Serrat, N. and Coenders, G. (2024), "Financial analysis of rural tourism in Catalonia and Galicia pre- and post COVID-19", *International Journal of Tourism Research*, Vol. 26 No. 4, Article e2698. https://doi.org/10.1002/jtr.2698

Seo, K. and Soh, J. (2019), "Asset-light business model: An examination of investment-cash flow sensitivities and return on invested capital", *International Journal of Hospitality Management*, Vol. 78, pp. 169-178. https://doi.org/10.1016/j.ijhm.2018.12.003

Seo, K., Woo, L., Mun, S.G. and Soh, J. (2021), "The asset-light business model and firm performance in complex and dynamic environments: The dynamic capabilities view", *Tourism Management*, Vol. 85, Article 104311. https://doi.org/10.1016/j.tourman.2021.104311

Spencer, D.M. and Holecek, F. (2007), "Basic characteristics of the fall tourism market", *Tourism Management*, Vol. 28 No. 2, pp. 491-504. https://doi.org/10.1016/j.tourman.2006.03.005

Strouhal, J., Stamfestová, P., Kljucnikov, A. and Vincúrová, Z. (2018), "Different approaches to the EBIT construction and their impact on corporate financial performance based on the return on assets: Some evidence from Czech Top100 companies", *Journal of Competitiveness*, Vol. 10 No. 1, pp. 144-154. DOI: 10.7441/joc.2018.01.09

Xiao, Q., O'Neill, J.W. and Mattila, A.S. (2012), "The role of hotel owners: the influence of corporate strategies on hotel performance", *International Journal of Contemporary Hospitality Management*, Vol. 24 No. 1, pp. 122-139. https://doi.org/10.1108/09596111211197836